\documentclass[submission,copyright,creativecommons]{eptcs}
\providecommand{\event}{TTC 2011} 

\usepackage{multirow}
\usepackage{graphicx}
\usepackage{subfigure}
\usepackage{url}
\usepackage{amssymb}
\usepackage{xspace}

\usepackage{amsmath}
\usepackage{listings}

\makeindex

\lstdefinelanguage{groovy} {
    emph={println, new, tokenize, each, def, static, for, if, else, in, assert},
    emphstyle=\bfseries,
    morecomment=[l]{//},
    basicstyle=\fontfamily{pcr}\scriptsize,
    string=[b]",
    showstringspaces=false
}

\newcommand{\operation}[1]{\textsf{#1}}

\newcommand{\myparagraph}[1]{\vspace{.5em}\par\noindent\textbf{#1}}

\title{Solving the TTC 2011 Model Migration Case with Edapt}
\author{Markus Herrmannsdoerfer
\institute{Institut f\"ur Informatik,
  Technische Universit\"{a}t M\"{u}nchen 
}
\email{herrmama@in.tum.de}
}
\def\titlerunning{Solving the TTC 2011 Model Migration Case with Edapt}
\def\authorrunning{M. Herrmannsdoerfer}
\begin{document}
\maketitle

\begin{abstract}
This paper gives an overview of the Edapt solution to the GMF model migration case \cite{modelmigrationcase} of the Transformation Tool Contest 2011.
\end{abstract}

\section{Edapt in a Nutshell}

Edapt\footnote{\url{http://www.eclipse.org/edapt}} is a transformation tool tailored for the migration of models in response to metamodel adaptation.
Edapt is an official Eclipse tool derived from the research prototype COPE.

\begin{figure}[htb]
\centering
		\includegraphics[scale=0.6]{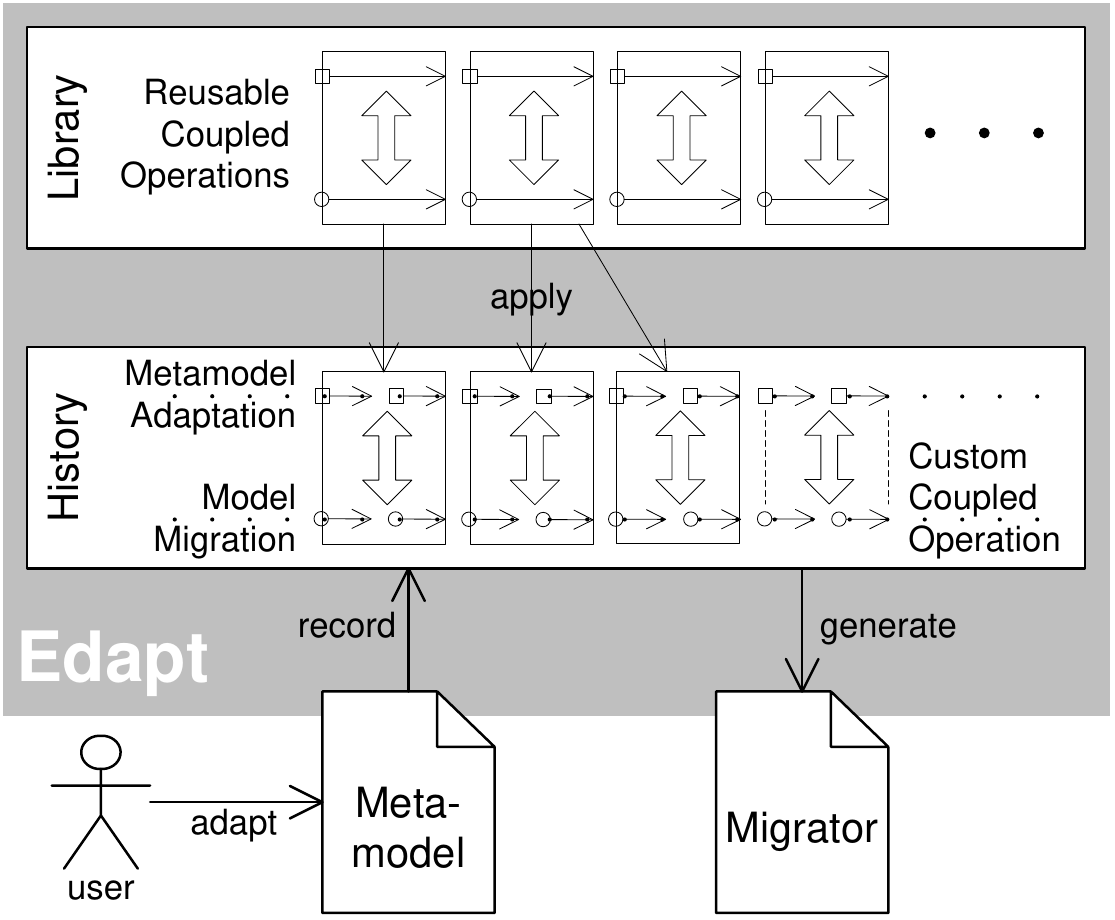}
				\vskip -5pt
\caption{Overview of Edapt}
\label{fig:overview}
\end{figure}

\myparagraph{Modeling the Coupled Evolution.}
As depicted by Figure~\ref{fig:overview}, Edapt specifies the me\-ta\-mo\-del adaptation as a sequence of operations in an explicit history model.
The operations can be enriched with instructions for model migration to form so-called coupled operations.
Edapt provides two kinds of coupled operations according to the automatability of the model migration~\cite{Herrmannsdoerfer2009_COPE-AutomatingCoupledEvolutionofMetamodelsandModels}: reusable and custom coupled operations.

Reuse of recurring migration specifications allows to reduce the effort associated with building a model migration~\cite{Herrmannsdoerfer2008_AutomatabilityofCoupledEvolutionofMetamodelsandModelsinPractice}.
Edapt thus provides \emph{reusable coupled operations} which make metamodel adaptation and model migration independent of the specific metamodel through parameters and constraints restricting the applicability of the operation.
An example for a reusable coupled operation is \emph{Enumeration to Sub Classes} which replaces an enumeration attribute with subclasses for each literal of the enumeration.
Currently, Edapt comes with a library of over 60 reusable coupled operations~\cite{Herrmannsdoerfer2010_AnExtensiveCatalogofOperatorsfortheCoupledEvolutionofMetamodelsandModels}.
By means of studying real-life metamodel histories, we have shown that, in practice, most of the coupled evolution can be covered by reusable coupled operations~\cite{Herrmannsdoerfer2008_AutomatabilityofCoupledEvolutionofMetamodelsandModelsinPractice,Herrmannsdoerfer2010_LanguageEvolutioninPracticeTheHistoryofGMF}.

Migration specifications can become so specific to a certain metamodel that reuse does not make sen\-se~\cite{Herrmannsdoerfer2008_AutomatabilityofCoupledEvolutionofMetamodelsandModelsinPractice}.
To express these complex migrations, Edapt allows the user to define a custom coupled operation by manually encoding a model migration for a metamodel adaptation in a Turing-complete language~\cite{Herrmannsdoerfer2008_COPEALanguagefortheCoupledEvolutionofMetamodelsandModels}.
By softening the conformance of the model to the metamodel within a coupled operation, both metamodel adaptation and model migration can be specified as in-place transformations, requiring only to specify the difference.
A transaction mechanism ensures conformance at the boundaries of the coupled operation.

\begin{figure}[!tb]
\centering
		\includegraphics[width=.85\textwidth]{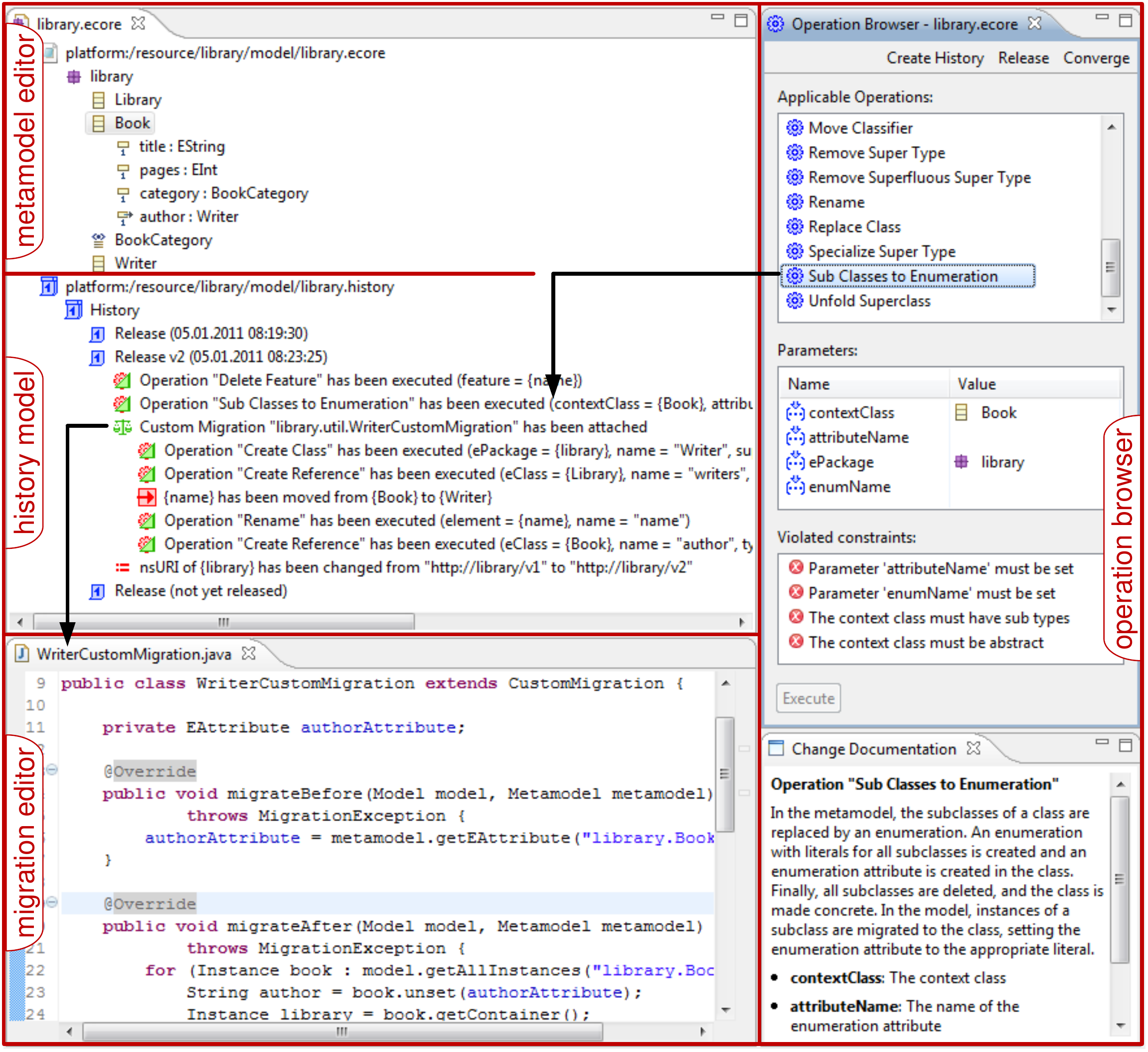}
		\vskip -5pt
\caption{User interface of Edapt}
\label{fig:screenshot}
\end{figure}

\myparagraph{Recording the Coupled Evolution.}
To not lose the intention behind the metamodel adaptation, Edapt is intended to be used already when adapting the metamodel.
Therefore, Edapt's user interface, which is depicted in Figure~\ref{fig:screenshot}, is directly integrated into the existing EMF \emph{metamodel editor}.
The user interface provides access to the \emph{history model} in which Edapt records the sequence of coupled operations.
An initial history can be created for an existing metamodel by invoking \emph{Create History} in the \emph{operation browser} which also allows the user to \emph{Release} the metamodel.

The user can adapt the metamodel by applying reusable coupled operations through the \emph{operation browser}.
The operation browser allows to set the parameters of a reusable coupled operation, and gives feedback on the operation's applicability based on the constraints.
When a reusable coupled operation is executed, its application is automatically recorded in the history model.
Figure~\ref{fig:screenshot} shows the operation \operation{Sub Classes to Enumeration} being selected in the operation browser and recorded to the history model.

The user needs to perform a custom coupled operation only, in case no reusable coupled operation is available for the change at hand.
First, the metamodel is directly adapted in the metamodel editor, in response to which the changes are automatically recorded in the history.
A migration can later be attached to the sequence of metamodel changes.
Figure~\ref{fig:screenshot} shows the \emph{migration editor} to encode the custom migration in Java.

\section{GMF Model Migration Case}

The complete solution is available through a SHARE demo \cite{modelmigrationsolutionedaptshare} and in the repository of the Eclipse Edapt project\footnote{\url{http://dev.eclipse.org/svnroot/modeling/org.eclipse.emft.edapt/trunk/examples/ttc_gmf}}.
Here, we only briefly describe the main characteristics of the solution.

\myparagraph{Core Task.}
To build the history model for the GMF Graph metamodel, we took the intermediate metamodel versions from the case resources and applied coupled operations to get from one metamodel version to the next.
To know which coupled operations to apply, we used the difference model obtained with EMF Compare\footnote{\url{http://www.eclipse.org/emf/compare/}} between two subsequent metamodel versions.
Edapt provides the so-called \emph{Convergence View} which automatically updates the difference model after each application of an operation~\cite{Herrmannsdoerfer2010_COPE-AWorkbenchfortheCoupledEvolutionofMetamodelsandModels}.

Table~\ref{tab:gmfgraph} lists the employed coupled operations together with their number of applications.
Most of the evolution can be covered by reusable coupled operations.
For more information about the reusable coupled operations, we refer the reader to \cite{Herrmannsdoerfer2010_AnExtensiveCatalogofOperatorsfortheCoupledEvolutionofMetamodelsandModels}.
Section~\ref{fig:gmfgraph} of the appendix shows the screenshot of the history model from release 1.0 to 2.1.
The history model also contains markers for the different metamodel versions.
Only two custom coupled operations are necessary to specify a correct model migration.
The first custom migration is necessary to initialize a reference that was made mandatory.
The second custom migration decouples the diagram elements from the figures to make the figures reusable.
The custom migrations which are implemented in Java can be found in Sections \ref{customone} and \ref{customtwo} of the appendix of this paper.

\begin{table}[tb]
\footnotesize
	\caption{GMF Graph}
	\label{tab:gmfgraph}
	\centering
		\begin{tabular}{|l|l|r|}\hline
		\bf Operation & \bf Kind & \bf Number \\\hline
		Add Super Type & Reusable & 10 \\
		Change Namespace URI & Reusable & 1 \\
		Create Attribute & Reusable & 2 \\
		Create Class & Reusable & 4 \\
		Create Reference & Reusable & 5 \\
		Delete Feature & Reusable & 4 \\
		Delete Operation & Reusable & 1 \\
		Document Metamodel Element & Reusable & 12 \\
		Drop Attribute Identifier & Reusable & 1 \\
		Extract Super Class & Reusable & 3 \\
		Generalize Attribute & Reusable & 1 \\
		Generalize Reference & Reusable & 1 \\
		Inline Super Class & Reusable & 2 \\
		Make Class Abstract when Interface & Reusable & 14 \\
		Make Reference Containment & Reusable & 1 \\
		Not Changeable to Suppressed Set Visibility & Reusable & 1 \\
		Push Down Feature & Reusable & 4 \\
		Remove Super Type & Reusable & 10 \\
		Specialize Reference Type & Reusable & 4 \\
		Specialize Super Type & Reusable & 6 \\
		Suppressed Set Visibility to Not Changeable & Reusable & 1 \\
		Unfold Super Class & Reusable & 1 \\
		Initialize FigureAccessor.typedFigure & Custom & 1 \\
		Decouple FigureHandle.referencingElements & Custom & 1 \\\hline
		\end{tabular}
\end{table}

\myparagraph{Multi-File Models.}
Edapt can migrate multi-file models, as it preserves the modularization into files throughout the migration.
If one file refers to the other, the second file is automatically loaded when resolving references.
However, a user can also explicitly define that several files are migrated together.

\myparagraph{GMF Map metamodel.}
To build the history model for the GMF Map metamodel, we applied the same procedure as in case of the GMF Graph metamodel.
Table~\ref{tab:gmfmap} lists the employed coupled operations together with their number of applications.
The complete coupled evolution can be covered by reusable coupled operations.
For more information about the reusable coupled operations, we refer the reader to \cite{Herrmannsdoerfer2010_AnExtensiveCatalogofOperatorsfortheCoupledEvolutionofMetamodelsandModels}.
The screenshots of the resulting history model are depicted in Section~\ref{fig:gmfmap1} of the appendix:
One screenshot shows the coupled operations from release 1.0 to 2.0, and another one the operations from release 2.0 to 2.1.

As one can see, the history model is modularized into different releases.
Based on the namespace URI, Edapt automatically detects of which release a model is and applies the coupled operations from this release to the newest release.
Moreover, the history model can be recorded for metamodels that refer to other metamodels.
However, to be able to load the models, the other metamodels need to be available during migration.
In the solution, we ensured this by starting the migration in an Eclipse workbench where the other metamodels are available through plugins.

\begin{table}[tb]
\footnotesize
	\caption{GMF Map}
	\label{tab:gmfmap}
	\centering
		\begin{tabular}{|l|l|r|}\hline
		\bf Operation & \bf Kind & \bf Number \\\hline
		Change GMF Constraint & Reusable & 5 \\
		Change Namespace URI & Reusable & 3 \\
		Create Annotation & Reusable & 1 \\
		Create Attribute & Reusable & 3 \\
		Create Class & Reusable & 2 \\
		Create Enumeration & Reusable & 3 \\
		Create GMF Constraint & Reusable & 6 \\
		Create Reference & Reusable & 3 \\
		Delete Annotation & Reusable & 2 \\
		Document Metamodel Element & Reusable & 12 \\
		Extract Subclass & Reusable & 1 \\
		Extract Super Class & Reusable & 1 \\
		Inheritance to Delegation & Reusable & 1 \\
		Make Class Abstract when Interface & Reusable & 10 \\
		Make Feature Volatile & Reusable & 1 \\
		Move Annotation & Reusable & 1 \\
		Not Changeable to Suppressed Set Visibility & Reusable & 5 \\
		Push down Feature & Reusable & 2 \\
		Replace Enumeration & Reusable & 2 \\
		Suppressed Set Visibility to Not Changeable & Reusable & 5 \\\hline
		\end{tabular}
\end{table}

\section{Conclusion}

\myparagraph{Expressiveness.}
The language provided by Edapt is expressive enough to specify the migration.
Most of the migration can be covered by reusable coupled operations.
For more complex migrations, custom coupled operations can be specified in Java.
The API provided to specify the custom coupled operations turned out to be expressive enough to cover the complex migrations involved in the GMF Graph metamodel evolution.

\myparagraph{Correctness.}
To ensure correctness, we used the test models from the case resources.
We modified the history model until there are no differences between the migrated model and the expected model.
To obtain the differences, we used EMF Compare.
We ignored differences in the order of references, since we believe that the order is not important for the GMF Graph and Map models.

\myparagraph{Conciseness.}
The migration specification is rather concise, since most of the evolution can be covered by reusable coupled operations.
Only two custom coupled operations are required: one with 21 and another one with 90 lines of custom migration code.
Considering the complexity of the second migration, we consider the migration specification as concise.
However, conciseness could be improved by providing a DSL instead of a Java API to implement migrations.

\myparagraph{Maintainability.}
The history model is easy to understand, since it is modularized into rather small coupled operations.
Intermediate metamodel versions can be reconstructed from the history model, thus simulating the evolution.
Moreover, coupled operations can be undone to revert changes that have been performed erroneously, and custom migrations can be removed, edited and added to fix error-prone migrations.
Finally, existing history models can be extended by recording new coupled operations that are applied to the metamodel.


\bibliographystyle{eptcs}
\bibliography{literature}

\begin{thebibliography}{1}
\providecommand{\bibitemdeclare}[2]{}
\providecommand{\urlprefix}{Available at }
\providecommand{\url}[1]{\texttt{#1}}
\providecommand{\href}[2]{\texttt{#2}}
\providecommand{\urlalt}[2]{\href{#1}{#2}}
\providecommand{\doi}[1]{doi:\urlalt{http://dx.doi.org/#1}{#1}}
\providecommand{\bibinfo}[2]{#2}

\bibitemdeclare{inproceedings}{Herrmannsdoerfer2010_COPE-AWorkbenchfortheCoupl%
edEvolutionofMetamodelsandModels}
\bibitem{Herrmannsdoerfer2010_COPE-AWorkbenchfortheCoupledEvolutionofMetamodel%
sandModels}
\bibinfo{author}{Markus Herrmannsdoerfer} (\bibinfo{year}{2010}):
  \emph{\bibinfo{title}{{COPE} - A Workbench for the Coupled Evolution of
  Metamodels and Models}}.
\newblock In: {\sl \bibinfo{booktitle}{SLE '10}}.

\bibitemdeclare{inproceedings}{modelmigrationcase}
\bibitem{modelmigrationcase}
\bibinfo{author}{Markus Herrmannsdoerfer} (\bibinfo{year}{2011}):
  \emph{\bibinfo{title}{{GMF}: A Model Migration Case for the Transformation
  Tool Contest}}.
\newblock In \bibinfo{editor}{\bibinfo{editor}{{Van Gorp}}} et~al.
  \cite{ttc2011eptcs}.

\bibitemdeclare{inproceedings}{modelmigrationsolutionedaptshare}
\bibitem{modelmigrationsolutionedaptshare}
\bibinfo{author}{Markus Herrmannsdoerfer} (\bibinfo{year}{2011}):
  \emph{\bibinfo{title}{{SHARE} demo related to the paper {Solving the TTC 2011
  Model Migration Case with Edapt}}}.
\newblock In \bibinfo{editor}{\bibinfo{editor}{{Van Gorp}}} et~al.
  \cite{ttc2011eptcs}.
\newblock
  \urlprefix\url{http://is.ieis.tue.nl/staff/pvgorp/share/?page=ConfigureNewSe%
ssion&vdi=XP-TUe_TTC11_EMFEdapt.vdi}.

\bibitemdeclare{inproceedings}{Herrmannsdoerfer2008_AutomatabilityofCoupledEvo%
lutionofMetamodelsandModelsinPractice}
\bibitem{Herrmannsdoerfer2008_AutomatabilityofCoupledEvolutionofMetamodelsandM%
odelsinPractice}
\bibinfo{author}{Markus Herrmannsdoerfer}, \bibinfo{author}{Sebastian Benz} \&
  \bibinfo{author}{Elmar Juergens} (\bibinfo{year}{2008}):
  \emph{\bibinfo{title}{Automatability of Coupled Evolution of Metamodels and
  Models in Practice}}.
\newblock In: {\sl \bibinfo{booktitle}{MoDELS '08}},
  \doi{10.1007/978-3-540-87875-9\_45}.

\bibitemdeclare{inproceedings}{Herrmannsdoerfer2008_COPEALanguagefortheCoupled%
EvolutionofMetamodelsandModels}
\bibitem{Herrmannsdoerfer2008_COPEALanguagefortheCoupledEvolutionofMetamodelsa%
ndModels}
\bibinfo{author}{Markus Herrmannsdoerfer}, \bibinfo{author}{Sebastian Benz} \&
  \bibinfo{author}{Elmar Juergens} (\bibinfo{year}{2008}):
  \emph{\bibinfo{title}{{COPE}: A Language for the Coupled Evolution of
  Metamodels and Models}}.
\newblock In: {\sl \bibinfo{booktitle}{MCCM '08}}.

\bibitemdeclare{inproceedings}{Herrmannsdoerfer2009_COPE-AutomatingCoupledEvol%
utionofMetamodelsandModels}
\bibitem{Herrmannsdoerfer2009_COPE-AutomatingCoupledEvolutionofMetamodelsandMo%
dels}
\bibinfo{author}{Markus Herrmannsdoerfer}, \bibinfo{author}{Sebastian Benz} \&
  \bibinfo{author}{Elmar Juergens} (\bibinfo{year}{2009}):
  \emph{\bibinfo{title}{{COPE} - Automating Coupled Evolution of Metamodels and
  Models}}.
\newblock In: {\sl \bibinfo{booktitle}{{ECOOP} '09}},
  \doi{10.1007/978-3-642-03013-0\_4}.

\bibitemdeclare{inproceedings}{Herrmannsdoerfer2010_LanguageEvolutioninPractic%
eTheHistoryofGMF}
\bibitem{Herrmannsdoerfer2010_LanguageEvolutioninPracticeTheHistoryofGMF}
\bibinfo{author}{Markus Herrmannsdoerfer}, \bibinfo{author}{Daniel Ratiu} \&
  \bibinfo{author}{Guido Wachsmuth} (\bibinfo{year}{2009}):
  \emph{\bibinfo{title}{Language Evolution in Practice: The History of {GMF}}}.
\newblock In: {\sl \bibinfo{booktitle}{SLE '09}},
  \doi{10.1007/978-3-642-12107-4\_3}.

\bibitemdeclare{inproceedings}{Herrmannsdoerfer2010_AnExtensiveCatalogofOperat%
orsfortheCoupledEvolutionofMetamodelsandModels}
\bibitem{Herrmannsdoerfer2010_AnExtensiveCatalogofOperatorsfortheCoupledEvolut%
ionofMetamodelsandModels}
\bibinfo{author}{Markus Herrmannsdoerfer}, \bibinfo{author}{Sander Vermolen} \&
  \bibinfo{author}{Guido Wachsmuth} (\bibinfo{year}{2010}):
  \emph{\bibinfo{title}{An Extensive Catalog of Operators for the Coupled
  Evolution of Metamodels and Models}}.
\newblock In: {\sl \bibinfo{booktitle}{SLE '10}},
  \doi{10.1007/978-3-642-19440-5\_10}.

\bibitemdeclare{proceedings}{ttc2011eptcs}
\bibitem{ttc2011eptcs}
\bibinfo{editor}{Pieter {Van Gorp}}, \bibinfo{editor}{Steffen Mazanek} \&
  \bibinfo{editor}{Louis Rose}, editors (\bibinfo{year}{2011}):
  \emph{\bibinfo{title}{TTC 2011: Fifth Transformation Tool Contest, Z\"urich,
  Switzerland, June 29-30 2011, Post-Proceedings}}.
  \bibinfo{publisher}{{EPTCS}}.

\end{thebibliography}

\newpage

\appendix

\section{Solution}

\subsection{GMF Graph History}

		\includegraphics[width=.75\textwidth]{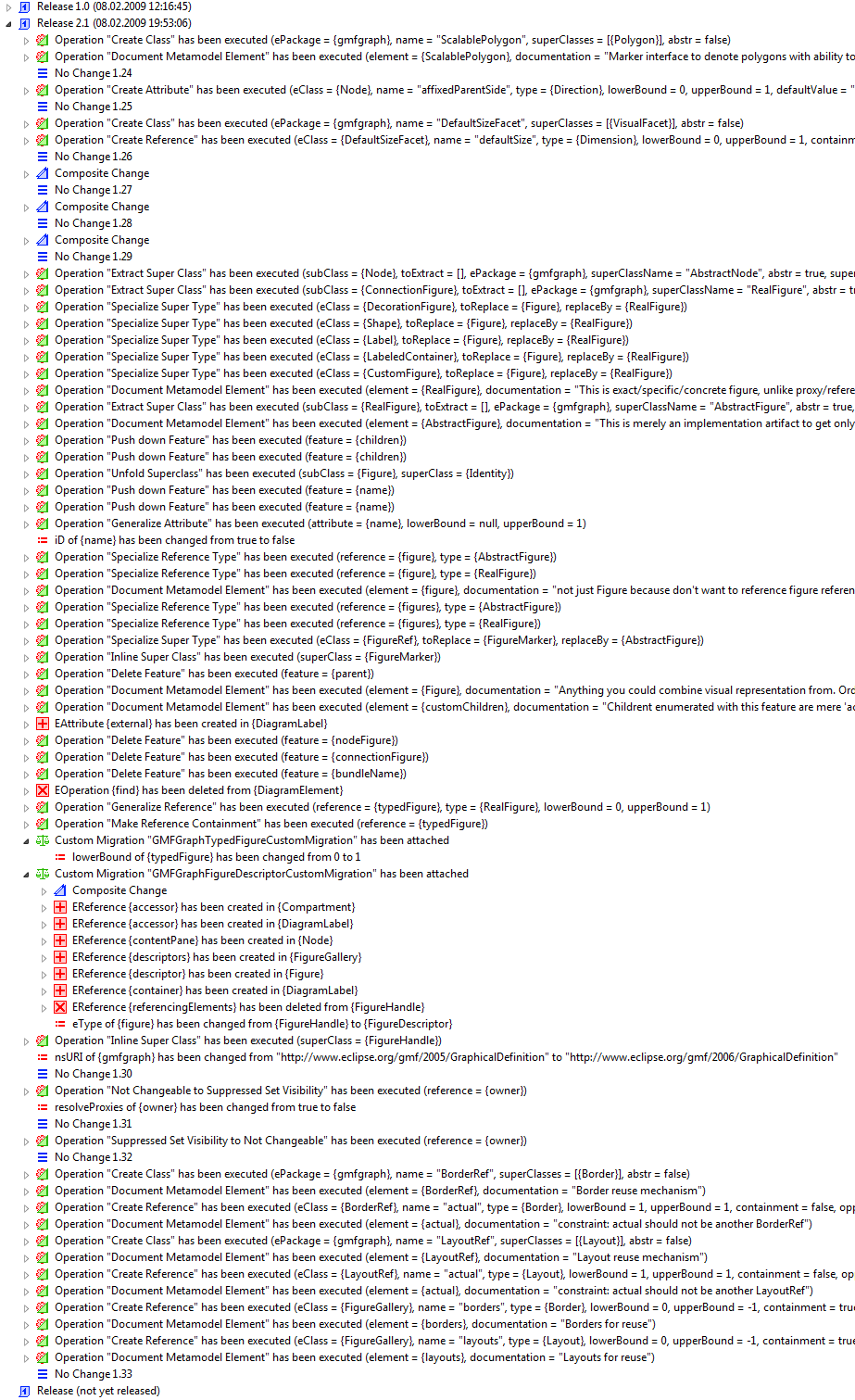}
	\label{fig:gmfgraph}

\subsection{GMF Map History}

		\includegraphics[width=.75\textwidth]{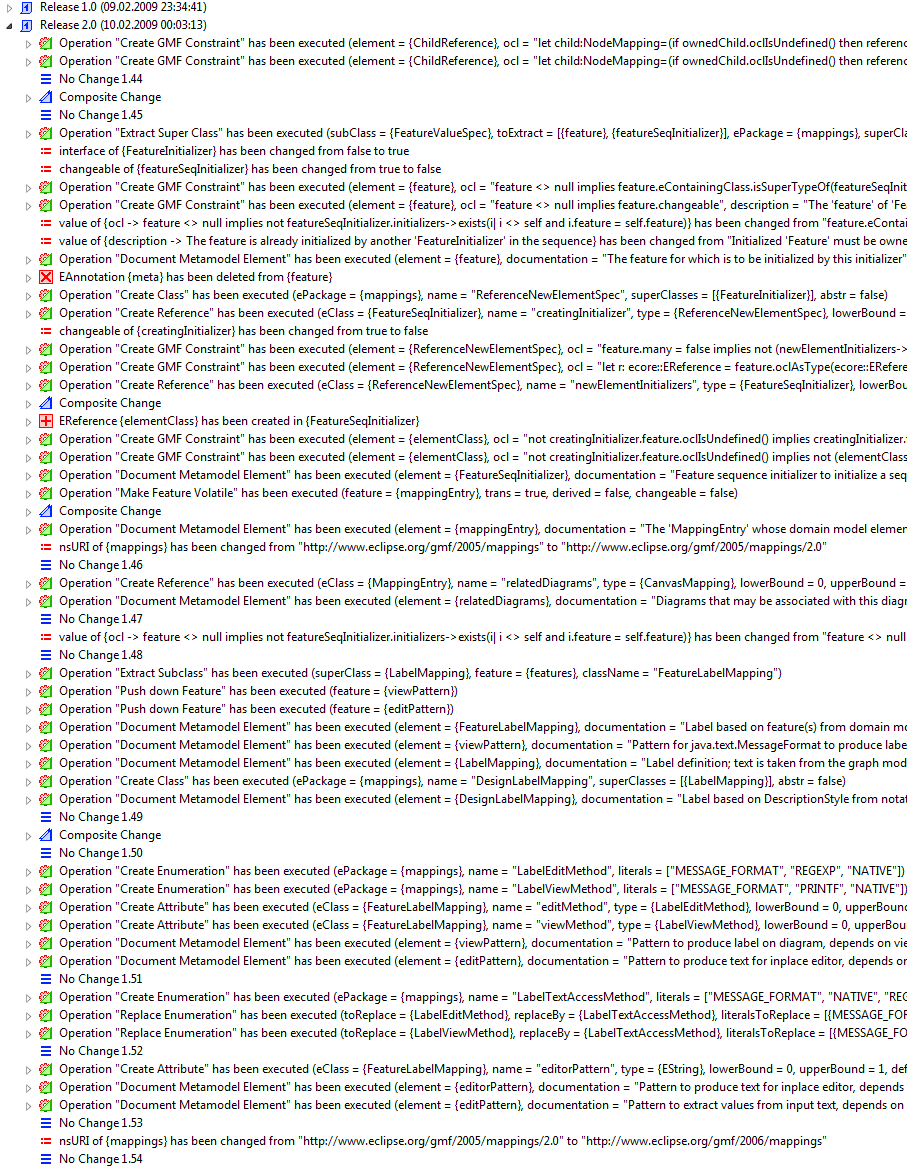}
	\label{fig:gmfmap1}

\noindent		\includegraphics[width=.75\textwidth]{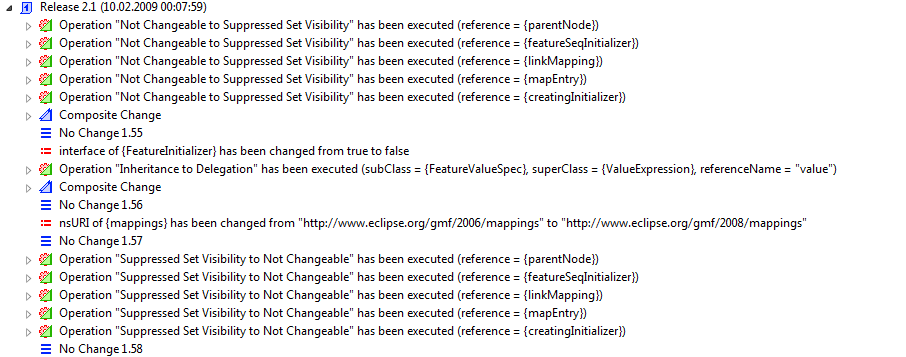}
	\label{fig:gmfmap2}

\subsection{Initialize FigureAccessor.typedFigure}\label{customone}
\begin{lstlisting}[language=java]
import org.eclipse.emf.edapt.migration.CustomMigration;
import org.eclipse.emf.edapt.migration.Instance;
import org.eclipse.emf.edapt.migration.Metamodel;
import org.eclipse.emf.edapt.migration.Model;

public class GMFGraphTypedFigureCustomMigration extends CustomMigration {

	@Override
	public void migrateAfter(Model model, Metamodel metamodel) {
		for(Instance fa : model.getAllInstances("gmfgraph.FigureAccessor")) {
			Instance tf = fa.getLink("typedFigure");
			if(tf == null) {
				tf = model.newInstance("gmfgraph.CustomFigure");
				tf.set("qualifiedClassName", "org.eclipse.draw2d.IFigure");
				fa.set("typedFigure", tf);
			} else {
				tf.set("name", null);
			}
		}
	}
}
\end{lstlisting}
\vspace{1em}

\subsection{Decouple FigureHandle.referencingElements}\label{customtwo}
\begin{lstlisting}[language=java]
import java.util.List;

import org.eclipse.emf.ecore.EReference;
import org.eclipse.emf.edapt.migration.CustomMigration;
import org.eclipse.emf.edapt.migration.Instance;
import org.eclipse.emf.edapt.migration.Metamodel;
import org.eclipse.emf.edapt.migration.MigrationException;
import org.eclipse.emf.edapt.migration.Model;

public class GMFGraphFigureDescriptorCustomMigration extends CustomMigration {

	private EReference reference;

	@Override
	public void migrateBefore(Model model, Metamodel metamodel)
			throws MigrationException {
		reference = metamodel
				.getEReference("gmfgraph.FigureHandle.referencingElements");
	}

	@Override
	public void migrateAfter(Model model, Metamodel metamodel)
			throws MigrationException {
		for (Instance handle : model.getAllInstances("gmfgraph.FigureHandle")) {
			List<Instance> elements = handle.unset(reference);
			if (!elements.isEmpty()) {
				Instance toplevel = getToplevel(handle);
				Instance descriptor = getOrCreateDescriptor(toplevel, model);
				for (Instance element : elements) {
					element.set("figure", descriptor);
				}
				if (toplevel != handle) {
					Instance access = getOrCreateAccess(descriptor, handle);
					for (Instance element : elements) {
						if (element.instanceOf("gmfgraph.DiagramLabel")
								|| element.instanceOf("gmfgraph.Compartment")) {
							element.set("accessor", access);
						}
					}
				}
			}
		}
	}

	public Instance getToplevel(Instance handle) {
		while (handle.getContainer().instanceOf("gmfgraph.FigureHandle")) {
			handle = handle.getContainer();
		}
		return handle;
	}

	public Instance getOrCreateDescriptor(Instance toplevel, Model model) {
		Instance gallery = toplevel.getContainer();
		if (gallery.instanceOf("gmfgraph.FigureDescriptor")) {
			return gallery;
		}
		Instance descriptor = model.newInstance("gmfgraph.FigureDescriptor");
		descriptor.set("actualFigure", toplevel);
		gallery.remove("figures", toplevel);
		gallery.add("descriptors", descriptor);
		descriptor.set("name", toplevel.get("name"));
		return descriptor;
	}

	public Instance getOrCreateAccess(Instance descriptor, Instance handle) {
		Instance figure = null;
		if (handle.instanceOf("gmfgraph.FigureAccessor")) {
			figure = handle.getLink("typedFigure");
		} else {
			figure = handle;
		}
		Instance access = findAccess(descriptor, figure);
		if (access == null) {
			access = descriptor.getType().getModel()
					.newInstance("gmfgraph.ChildAccess");
			access.set("figure", figure);
			descriptor.add("accessors", access);
		}
		return access;
	}

	public Instance findAccess(Instance descriptor, Instance figure) {
		for (Instance access : descriptor.getLinks("accessors")) {
			if (access.get("figure") == figure) {
				return access;
			}
		}
		return null;
	}
}
\end{lstlisting}
\vspace{1em}

\end{document}